\documentclass{article}
\usepackage{fullpage}

\usepackage{comment}

\usepackage{placeins}

\usepackage{amsmath,amssymb}

\usepackage[utf8x]{inputenc}

\usepackage{textcomp,marvosym}

\usepackage{hyperref}

\usepackage{nameref,hyperref}

\usepackage[table]{xcolor}

\usepackage{array}

\usepackage{lastpage,fancyhdr,graphicx}

\usepackage[frozencache=true,cachedir=minted-cache]{minted}

\author{Andy Somogyi, Herbert Sauro, James Glazier }
\date{December 2020}

\usepackage{graphicx}

\newcommand{\inpython}[1]{\mintinline{python}{#1}}

\begin{document}

\vspace*{0.2in}

\begin{flushleft}
{\Large
\textbf\newline{Real-Time Interactive Modeling and Simulation in Biological Physics and Active Matter with Mechanica} 
}
\newline
\\
Endre T. Somogyi\textsuperscript{1\ddag*},
Jeffery Coulter\textsuperscript{1},
Fanbo Sun\textsuperscript{1},
Herbert M. Sauro\textsuperscript{2},
James A. Glazier\textsuperscript{1},
\\
\bigskip
\textbf{1} Department of Intelligent Systems Engineering, Indiana University, Bloomington, IN, US 
\\
\textbf{2} Department of Bioengineering, University of Washington, Seattle, WA, US
\bigskip

%
%
\ddag Primary developer

* andy.somogyi@gmail.com

\end{flushleft}

\section*{Abstract}
Modeling and simulation (M\&S) has revolutionized the way modern engineered products are designed, tested and evaluated. Yet modeling and simulation is much less frequently applied in the study of natural biological and active matter systems. Two of the greatest challenges in applying M\&S approaches to natural biological systems are (1) difficulty in specifying a model, and developing a simulation from it, and (2) tuning and optimizing parameters. Here we address the first challenge directly, and develop a software library that can lead to progress in the second. 


\section{Introduction}


Interactive simulation is likely to be key to increasing scientific productivity. In this article we wish to present Mechanica, an \emph{Active Matter} simulation platform, built on an interactive, lattice-free numerical simulation engine with a consistent modeling formalism that enables users in a wide range physical science disciplines to experiment with and analyze simulations typical of the user’s discipline.  Mechanica is  designed  first  and  foremost  to  enable  users  to  work  interactively  with  simulations  –  so they can build a virtual environment, run a simulation in real-time, and interact with that simulation whilst it’s running. Mechanica is designed to be embedded into existing applications, and directly usable from Python as well as environments such as Jupyter notebooks. The core of Mechanica a native compiled C++ shared library, built on task-based parallelism, with an optimized memory layout. 

Active matter are physical systems which can be described as a large number of interacting elements which (1) have local control, (2) consume energy and move in their environment by exerting forces directly on their environment or other elements, and (3) are out of thermal equilibrium. A very wide range of physical systems (described in detail in section~\ref{sec:problem}) can be defined as active matter systems. 

Modeling and Simulation tools excel at representing engineered systems such as mechanical assemblies, fluid flows with fixed geometries, circuits, chemical reactions and molecular dynamics. However, the challenge with biological and active matter is these tend to tightly couple mechanical and chemical processes, and spatial relationships and agent identity continually change~\cite{Brodland:2015bl}.

Modeling,  simulation  and  analysis  platforms  such  as  SolidWorks,  Fusion360,  SPICE, Mathematica, Modelica, LabVIEW, and Simulink have revolutionized the way modern products  are  designed  and  developed.   There  are  a  number  of  key  aspects  of  these  simulation environments that differentiate them from “just” computer programming languages.  
(1) Simulation environments enable model construction by assembling and configuring meaningful, domain-relevant primitives.  
(2) They enable users to run, interact with, and experiment with a model without requiring any programming expertise on the part of the user. 
(3) They can simulate and analyze models with little or no delay, as most simulation environments incorporate a ”read-simulate-display” loop where users can interactively explore models, much like  one  experiments  with  new  designs  in  a  physical  research  laboratory.   

\subsection{Problem Domain}
\label{sec:problem}
Conceptualizing active matter physical phenomena is typically much harder than conceptualizing engineering problems because
 biological phenomena generally combine chemical and mechanical processes, whereas engineers model in either a chemical or a mechanical environment. Control in a biophysical environment is often a mix of chemical and mechanical decision-making.  In this realm, agents often change, both mechanically and chemically, the elements they interact with, and
 when control systems are designed  for engineered systems, the logic is typically encoded in a computer programming language.  These languages are designed for the purpose of codifying, encoding and exchanging decision making algorithms.  
 
\begin{figure}[h!]
\centering
\includegraphics[width=0.9\textwidth]{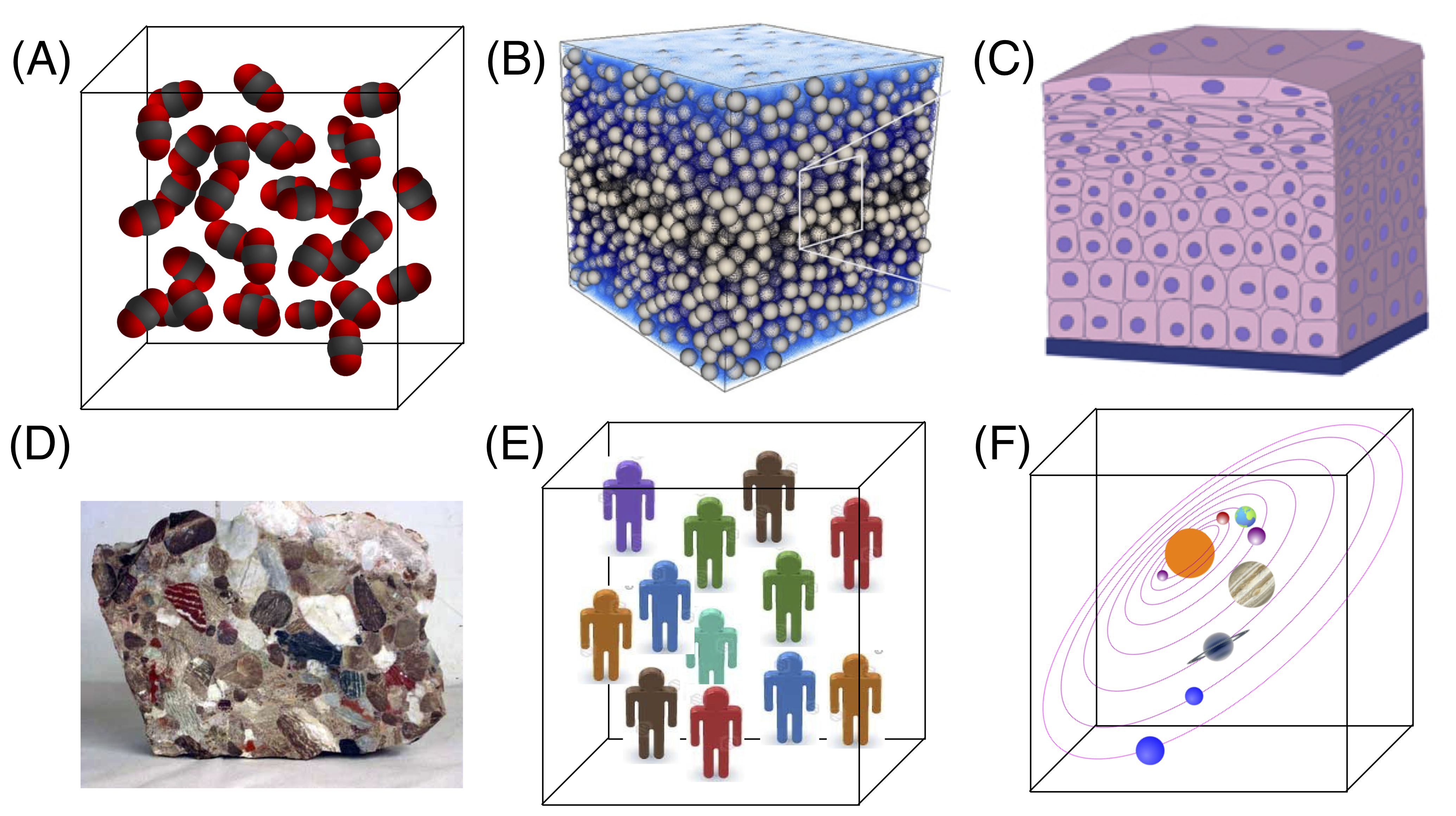}
\caption{Lets examine a number of naturally occurring systems, (A) is a liquid or gas composed of simple molecules, (B) is a fluid composed of small fluid "parcels", (C) is a section of biological cells in a tissue, (D) is a stone or mineral, a conglomerate, (D) is a group of individuals in a crowd, and (F) is a group of planets in a solar system. In each case, we can observe that each system is composed of a number of interacting objects or \emph{agents}. Some of these agents are simple, some are complex and have complex hierarchical structure and local control. We are developing a modeling framework that can handle all of these cases in a consistent way. }
\label{fig:things}
\end{figure}

In biological systems, we rarely have this luxury, and have to resort to encoding decision making in chemical networks. One of the greatest challenges at the  medium length scales found in chemical networks is that these dynamics and behaviors typically cannot be derived from first principles.  Rather, the observed behaviors are described phenomenologically or empirically.  Thus, the scientist exploring these kinds of phenomena needs a great deal of flexibility to propose and experiment with different kinds of interactions.
To demonstrate this difference, Figure~\ref{fig:things} represents a range of what seem to be completely disparate systems.  In (A) we draw a box around a collection of individual atoms and molecules.  These could be a protein, ligand, or any other classical molecular system.  In (B) we draw our box around a region of fluid.  (For example, we look at around 1+ micron range, with a moderate Reynolds number.)  In (C) we look at a similar sized region of space, but focus on  a  collection  of  biological  cells  in  an  organized  tissue such as  epithelium,  or  a  less organized collection of cells such as a group of cancerous cells in a tumor.  Then in (D) we look at a region of complex solid material, such as a stone aggregate, or metal alloy.  In (E), we look at what may seem like a completely different system:  a biological population where some members might be infected with a dangerous pathogen.  Finally, in (F) we step even further back, and look at a planetary system.
All of these system, and in fact the vast majority of any realizable physical system the authors are aware of, admit a similar description, and can be described in terms of objects and processes. In the most abstract sense, objects are stateful things that in and of themselves do not change. Processes act on objects to change their state. For our purposes here, we use a slightly more specific description, and we define our most basic object as a well-defined region of space that occupies a volume, and has a position, has an identity, and is an instance of a specific type. We define this object as a particle. A particle can also contain other smaller objects. Sometimes we might not be  concerned with the internal structural and spatial organization of the agents we are modeling, and all we are concerned about is that a region of space contains some amount of other substances, such as in a fluid. Other times, we might be concerned with the internal structure. But nonetheless this same description is, we believe, fairly universal.

\begin{figure}[h]
\centering
\includegraphics[width=0.7\textwidth]{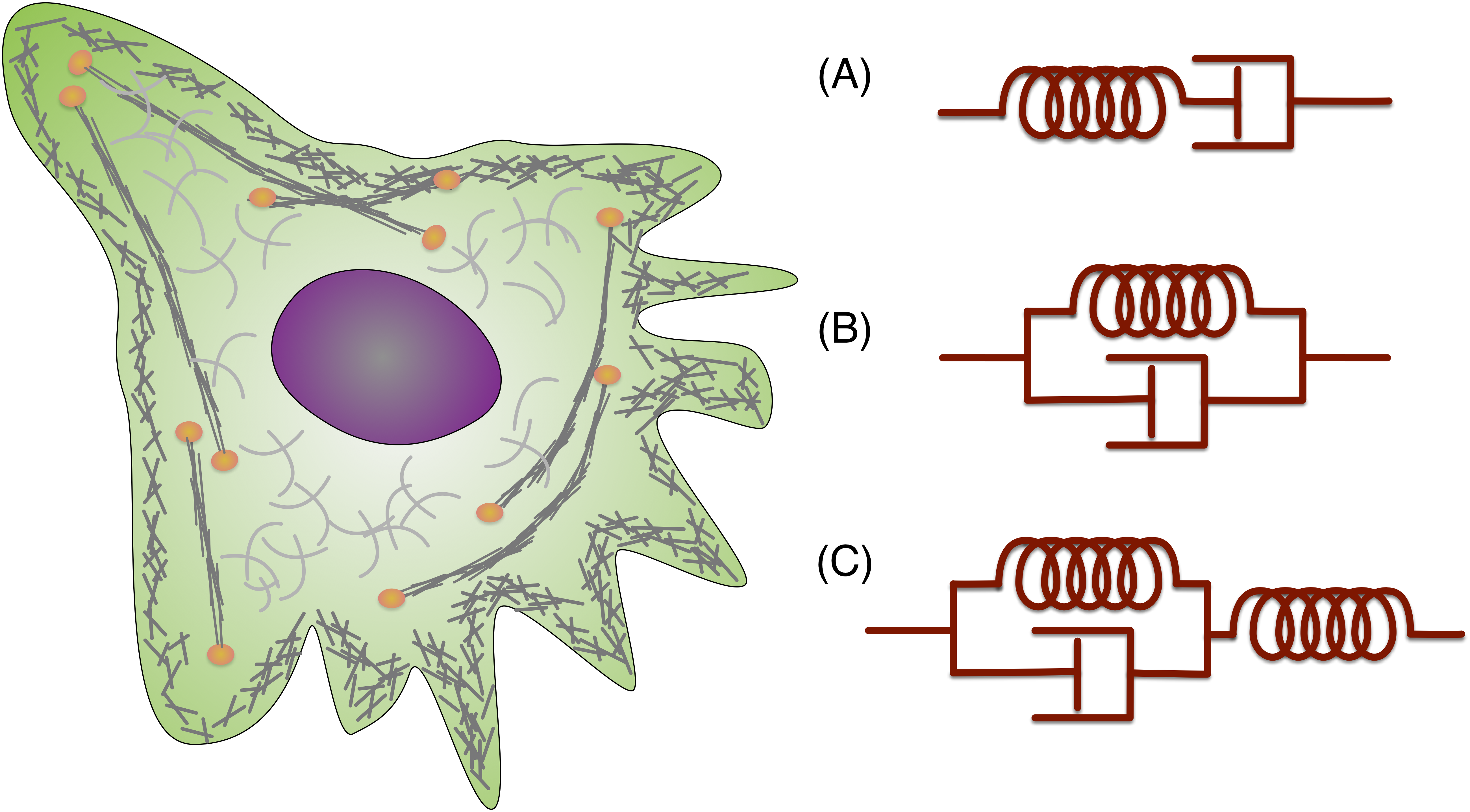}
\caption{A biological cell contains a very wide range of mechanical mechanisms, and most of these mechanisms permit an equally wide range of conceptualizations. For example, we might want to treat the mechanical behavior of actin filaments as either (A) a Maxwell, (B) a Kelvin, or (C) a Kelvin-Voigt model, or some other model. The choice of model to represent these physical systems depends on the problem at hand, and users need to be able to pick and choose from a \emph{toolbox} of components that they can assemble into larger, more complex models. }
\label{fig:cell}
\end{figure}

In these systems, how we define the exact regions varies from system to system. In the atomic example (A), the choice is obvious: we can refer to each individual atom in the system as a particle. In this most basic example, in the classical interpretation, atoms are atomic, they have no internal structure, and are internally homogeneous. In more continuum examples, e.g., in a material (C) or a fluid, (B), the choice of where to draw our boundaries is more arbitrary. In a continuum such as these, we may equivalently draw little rectangular boxes, or hexagonal boxes, or Voronoi boxes and call these our particles. In these examples, as in fluids, or aggregates, we can see that each one of our spatial regions contains some amount of other stuff, typically we call these \emph{chemical solutes}. In a biological tissue, the boundaries are again much more well defined, as the definition of a biological cell is that it has a well-defined cell membrane -- a boundary that separates its internal contents from the outside world -- and thus a reasonable choice might be to treat biological cells as particles. But what if we want more detail, what if we cannot neglect the complex internal structure of a cell, as in Figure~\ref{fig:cell}.

Here, looking a little closer, we might want to represent the cell nucleus as a larger particle, the actin filaments as rod-like particles, and cytoplasm as fluid-like particles. 
Going up a length scale, let's consider a small population of humans, where each of them carries some pathogen. Here we might choose to represent each individual as a particle, and each agent carries along with them some amount of materials as they move about and interact in space. Moving farther out, let’s take a look at a planetary system. The choice of how to delineate particles again becomes clear, as we can represent planets as particles. Here each planet itself is a well defined region of space and carries a very complex amount of stuff along with it. 
Now, let's consider how we can describe how these objects (particles) change, how they move around in space, how their local state can change, or in general, the dynamics of these systems. In this paper, we will examine how change in most physical systems can be described in terms of generalized forces, constraints and reactions. We define a generalized force as simply a propensity for an object to move in a particular direction. A constraint is a process that maintains a certain value or spatial relationship, and a reaction is a process that transforms materials (which can be either solutes attached to a particle or particles themselves) into material of a different type. 

Starting with our basic atomistic example, we can identify three basic classes of forces: bonded, non-bonded and external. We define a bonded force (or bonded interaction) as a persistent relationship between two or more object instances. For example, a chemical bond exists between individual atom instances, and is not directly affected by any external factors. A bonded interaction is very general. For example, say we have two biological cells, and a tight junction or some other mechanical bond exists between them. Or say we have instead individual actin molecules in a long actin filament where each one of these members is bonded to their neighbor with a bond. These bonds are a persistent relationship, and they themselves can have a local state and can evolve in time. For example, we might want to model viscous materials with a Kelvin-Voigt or Maxwell bond, where the bond rest length changes over time and might eventually break. 

Non-bonded interactions are interactions between types or families of objects, and  depend only on the types of objects and their spatial relationship. The most concrete examples of non-bonded or \emph{field} interactions are electro-static or gravitational, as in the atomistic example where electrostatic interaction  depends only on the charge of the individual objects and their spatial relationship. We can also represent fluid, biological cell, and planetary systems with non-bonded interactions. Fluid particles tend to move from areas of high to low pressure, and this pressure field is defined by the density of neighboring particles, very similar to a electro-static or gravitational repulsion. Non-bonded applies equally well to biological examples, such as biological cells that might feel a propensity to move in a particular chemical gradient (chemotaxis), or humans who might feel a propensity to move away from foul-smelling individuals. All of these are examples of a generalized non-bonded interaction. Constraints try to maintain a particular spatial relationship; for example, we might define a very rigid chain as being connected via a set of constraints instead of stiff bonds, or we might define a complex spatial boundary as having a volume constraint. Constraints are very similar to forces, but have some subtle differences. 

Reactions are the final class of transformative processes that we consider. A chemical reaction is probably the most familiar kind of transformative reaction. These transform from a set of local reactants into a set of local products. If model biological cells or parcels of fluid as individual particles, we can represent their local chemical dynamics as a set of reactions. But these local solutes have a tendency to spatially move from areas of high to low concentration. We define local solute reactions as simply reactions, and movement of solutes between object instances as flux processes. Reaction networks can model a wide range of cellular processes, such as gene regulatory, metabolic, signaling, inter alia.  Reactions can also describe the transformation of discrete objects; for example, a particle of a particular type could change to another type in response to some condition, or a particle instance might divide into two, such as in cellular mitosis, or die. 
We have covered a wide range of physical systems here, and shown how they can all be described in terms of processes acting on objects. The concepts of generalized forces, constraints, reactions, hierarchical particles, and attached solutes are extremely general, and are not tied to any particular problem domain or length scale. They are also quite abstract, and not tied to any particular programming language. In fact, we could implement these concepts with most object-oriented programming languages. In the later sections of this paper, we describe how Mechanica’s Python API enables users to build and simulate a number of models using these concepts, and how we have developed a native Python API that maps these general physical concepts into very natural Python language concepts.

This presents a significant challenge, as simulation environments that make it simple for the end user to write models without resorting to hard-coding C++ or FORTRAN are usually very limited in the level of flexibility they provide the end user. For example, if users want to write a standard molecular dynamics model, there are many different, really good choices of simulation engines and these kinds of models can easily be specified by human readable configuration files. However, as the kinds of interactions are not well standardized or formalized at medium length scales, users almost always are forced to resort to hard-coding FORTRAN or C++.

We will begin by first looking at a range seemingly disparate natural phenomena, then discuss how they can be described in terms of similar mechanistic building blocks, and end by discussing how we have created computational modeling constructs that implement these core mechanisms. 

When we use a mathematical or computational abstraction to represent physical systems, we need to always take into account that whilst the universe (probably) has infinite precision, we have a finite amount of detail we can represent in a computer. A computer model will always be an abstraction or idealization of a physical concept. We have a computational budget, there is a limit to how many concurrent things we can represent on a computer. When we model physical systems, we must choose wisely when spending our computational budget. The choice of computational representation is often dictated by the length scale under study. However, there are very many common concepts and mechanisms that are length scale invariant. In this case, we first enumerate a variety of different physical systems and discuss their common mechanisms, before discussing the computational constructs Mechanica provides that implement these mechanisms. 

Mechanica is a native compiled C++ shared library with a native and extensive Python API, that’s designed to be used from an ipython console (or via scripts, of course).

\subsection{Design Goals}

We established four principle design goals in developing Mechanica: (1) modeling capabilities, (2) real-time interactive simulation, and (3) software and component re-use and interoperability. 

The combination of a uniform and consistent representation of physical mechanisms that implement observed biological behaviors, a simple easy to use modeling API, real-time interactive 3D output, and the fact that Mechanica is a self-contained embeddable software library differentiate this package. 

\paragraph{Modeling Capabilities} 
The first and foremost requirement of any simulation software is the ability to actually represent, model, simulate and analyze things in the problem domain. In previous papers, we have made a series of studies in what are the core \emph{mechanisms} in active matter and biological physics~\cite{Somogyi:2016em, Somogyi:2017ws, Somogyi:2016be, Somogyi:2016kt, somogyi2018dynamic}. We have identified the core physical concepts or \emph{mechanisms} that we believe can serve as a \emph{basis set} of fundamental building blocks with which to construct biological and active matter simulations. Mechanica is built directly on the basic physical concepts, and treats them in a uniform and consistent way.


\paragraph{Real-time interactive simulation}
One of the greatest productivity multipliers of engineering modeling application such as SolidWorks is the ability to quickly build and model, and simulate it, in real-time. Users can manipulate, interact with, and test and fit a model, and use these programs almost as a "virtual" laboratory where they can experiment with and try out new ideas with instant feedback. 

Most interpreted programming languages such as Python, Scheme, Mathematica, MATLAB also enable real-time experimentation, feedback and program exploration as they operate in what's called read-eval-print (REPL) loop, where the user enters expressions in a command prompt, or notebook, and the language run-time instantly evaluates that expression and returns the result. This allows the user to rapidly explore and experiment with new concepts, and dynamically change their code structure. This is in contrast to compiled languages such as C++ which (1) have orders of magnitude more complexity in terms of language and syntax structure, and (2) must be compile and run, and programs can only be run as a whole. This means that a program typically can't be modified as it's being run, and makes experimentation more difficult. 

Mechanica seeks to combine the real-time 3D simulation and visualization capabilities commonly found in engineering simulation software, with the REPL concept of interacting with interpreted programming languages. 

This does present a bit of a challenge, when one writes a complete simulation in some programming language, we must write a simulation loop, which advanced the model forward in time, and displays output at each step. The challenge here is how can we incorporate a REPL whilst a simulation is running. We address this challenge in \ref{sec:arch} where we describe how we enable a REPL whilst a model is executing.

\paragraph{Software and component re-use and interoperability} 

We approach inseparability on several levels. (1) Mechanica is a self-contained \emph{library} with both C and Python APIs, and is fundamentally designed to be hosted and integrated into different applications. (2) Users write models in standard Python, and we've structured our modeling formalism to structure and simplify creating model components and packaging them into different python packages. (3) Since our modeling API is Python, any Mechanica model can be directly used with any Python packages. At the binary level, most of our Python methods return or accept standard Numpy arrays~(\url{https://numpy.org/}), this makes is easy to exchange data without converting it. (4) Visualization is a major goal with Mechanica, however we have gone to great lengths to remain not only cross-platform operating system wise, but we are not locked into any particular toolkit. For example, many simulation packages use the "QT" toolkit. This is a very large operating system abstraction layer combined with graphical user interface toolkit, which emulates the look and feel of the native operating system. We instead do not use any toolkit, but rather draw directly into an OpenGL frame buffer which can be hosted in \emph{native} windowing system toolkits. Furthermore, we also support off-screen rendering, with no windowing system whatsoever. This enables us to host Mechanica in web server environments such as Jupyter, do multi-process rendering, such that Mechanica can render in one process to an off-screen frame-buffer, and copy those bits via a message to another process that can blit them to a native window. This is the identical rendering process all modern web browsers implement. 

To address model interoperability and extensibility, we have developed a modular model specification formalism (schema), and use standard Python to construct and script models. At first, it may seem counter-intuitive that Python, a language widely known for it's slow run-time performance enables a high-performance simulation environment. However, Mechanica itself is written in C++, and we implement the Python modelling API in C++. This multi-language solution enables us to have a large separation of model specification from internal computational implementation.  We leverage the strengths of Python, namely, it's simplicity, ease of use, interactivity and rapid development, whilst at the same time leveraging the advantages of low-level languages like C++ to make the most use of the available hardware. 

Many simulation environments environments allow model specification only with configuration data files such as most molecular dynamics packages. This approach may make model interchange simpler (provided the file formats are well-documented), it also greatly limits end-user flexibility.  Some simulation packages do model specification in C++. This places a very large cognitive burden on the end-user, in the end-user, typically a physical scientist is forced to learn a complex programming language like C++, but also has to install compilers, and set up software build systems. 

Model specification in C++ can also place limits on performance, in that a high-performance software architecture tends to be very difficult to understand, especially when dealing with the nuances of processor architecture, threading, etc... We have a large separation between computational implementation and end-user model specification. This allows us a great deal of flexibility to change internal software architecture without effecting end-user model specification, and allows us to leverage processor vectorization,  multi-threading, and GPUs.

\section{Related Work}

There are a wide range of simulation software platforms that are focused towards particular physical length scales, or specific scientific disciplines, and some that are more general simulation frameworks. 

There are a number of different \emph{conceptualizations}, or \emph{descriptions} we can make of physical reality, that is, what kind of mathematical or computational constructs we choose to describe physical phenomena with. There is no single best approach, but some approaches may be better for certain kinds of physical phenomena that display certain kinds of structure or behaviors. 

Current simulation tools include both ODE-based (for modeling reaction kinetics, signalling, etc.)~\cite{Hoops:2006ui, Sauro:01, Somogyi:2015iz}  as well as agent-based virtual tissue tools (cells, tissues, organs) such as CompuCell3D~\cite{Swat:2012iz}, PhysiCell~\cite{Ghaffarizadeh:2018de}, Chaste~\cite{Mirams:2013it}, or Morpheus~\cite{jyothi2016morpheus} and a number for biologically oriented reaction-diffusion simulations such as Tissue Simulation Toolkit~\cite{daub2015cell}, Smoldyn~\cite{andrews2010detailed}, or VCell~\cite{Moraru:2008iv}, or MCell~\cite{kerr2008fast, stiles2001monte}. These can simulate spatial chemical processes such as reaction-diffusion using both stochastic particle based approaches such as MCell or Smoldyn, or Eulerian finite-volume such as V-Cell. MCell models are specified with
a natural and elegant rule-based approach similar to BioNetGen~\cite{Faeder:2005ud}. However, most of the non-agent-based
cell simulators can only model chemical processes with fixed geometries, and do not appear to model mechanical or electrical processes or dynamic morphology. MCell for example is a
particle based simulator, but it is restricted to point particles with no volume exclusion.

Most agent-based biological cell simulation tools such as CompuCell3D, Chaste or Morpheus tend to only support one simulation methodology, and are focused on a particular problem domain.  For example, CompuCell3D and Morpheus are built on the \emph{Cellular Potts} formalism, and only support such Eulerian lattice type models. Others are built on the \emph{Center Model} formalism, and only support simple point like cell particles. These tools tend to treat mechanics and chemical transport in different ways, which can be confusing to the user. Mechanica on the other hand supports range of different particle based modeling formalism that integrate chemical and mechanical processes in a consistent way. 

Cell simulation tools also to be implemented as complete stand-alone programs, rather than \emph{libraries} as Mechanica is. This makes re-using them difficult, as they are not designed to be used \emph{inside} or \emph{plugged into} other software. This also complicated using them with parameter optimization tools, as parameter optimization is generally based on the idea of optimizing an objective function, where the objective function is a user provided function. CompuCell3D does support model construction with a mix of Python and XML, but many cell simulation tools require users to hard-code models in C++, much like general purpose particle simulation libraries such as LAMMPS~\cite{Plimpton:1995wl}.

For fluids / materials simulation, a number of environments choose the Lagrangian, or particle based fluids approach, wheres others choose an Eulerian meshed approach. The Lagrangian approach follows a ``parcel'' of material around, chunks of material move around and we track their position, momentum, and any associated state variables with it. We conceptualize a material by assuming the material to be composed of a large number of parcels, or particles, and follow their respective motion. We follow it as it moves, and monitor change in its properties. The properties may be velocity, temperature, density, mass, or concentration, etc in the flow field. The Eulerian approach partitions space into a series of explicit volumes, typically a regular lattice, and track the rate of change of material in these control volumes. This approach monitors  the properties of a material at a given point in space as a function of time. 

Highly dynamic, performance critical simulation environments such computer games or combustion research typically use Lagrangian methods, as these methods are much better suited to dynamically changing simulation elements and boundary conditions. With fixed geometry however, such as most engineered applications, solid mechanics, they typically use Eulerian approaches, as they only incur the cost of computing a mesh once at the start of the simulation, and with fixed geometries, Eulerian approaches can perform better as they can pre-comute and optimize a mesh ahead of time, which can reduce the number of mesh volumes. In summary, when the problem domain is dynamic, model components undergo significant re-organization, Lagrangian methods tend to work better, but with fixed or lower deformation geometries, Eulerian methods tend to work better. Mechanica is designed as an active matter simulation enviorment, which shares significant commonality with the real-time simulation as in computer games, thus we have chosen to use an Lagrangian numerical method.

A wide range of \emph{models} are constructed using general purpose particle dynamics engines such as  LAMMPS or HOOMD-blue~\cite{anderson2020hoomd}. LAMMPS in particular is very flexible, and users generally build derived simulation tools with C/C++. HOOMD-blue does support model construction using a Python API, however these libraries tend to be focused on large-scale parallel molecular dynamics simulation, don't support real-time visualization and are not really intended towards interactive modeling and experimentation. 

Classical molecular dynamics, where we conceptualize individual atoms as point particles are almost always performed by Lagrangian simulators, in fact this is the one of the proto-typical applications of this approach. There are a wide range of molecular dynamics (MD) simulation tools such as NAMD~\cite{phillips2005scalable} or GROMACS~\cite{gromacs:2021}, which are designed to read models written in a textual format. However, these tools tend to be highly focused on well-defined molecular models with a \emph{fixed} number of objects, and do not support particle creation, deletion, or bonded relationship rearrangement at run time, and they rarely support interactive use.

Interchange formats such as the Systems Biology Markup Language (SBML)~\url{http://www.sbml.org/}
can describe a restricted phenomenon such as chemical reactions in a well-stirred compartment, but
have no concept of forces or dynamic geometry or structural rearrangement, and are too restrictive
for use as general modeling languages. SBML models can be connected together as part of a larger
simulation environment such as discrete event simulation~\cite{Belloli:2016tu}.

\section{Approach}

This section will describe some of Mechanica's capabilities, and how we can use Mechanica to enable users to build and interact with models. Mechanica models are written in standard Python, and can operate either from a browser or as a desktop tool. 

At the base of Mechanica is the notion of an object. Objects can be things such as molecules, cells, membranes, ion channels, the extra-cellular matrix, fluids, etc. Objects have quantifiable characteristics such as location, amount, concentration, mass and volume. Objects define a state, and can inherit and extend other objects, or may contain other objects. Objects are grouped into two categories: continuous and discrete. Continuous objects describe things such as continuously valued chemical or molecular fields which have position-dependent concentrations. 

\subsection{Overview -- Equations of Motion}

Mechanica is based on the idea of enabling models at different length scales. Individual model components or \emph{agents} are not atomic point-like particles like in the molecular dynamics, but rather they are regions of space. We define this basic agent a \emph{particle}. These regions of space often move around and interact with other regions, and they carry along embedded, smaller objects with them, these are usually dissolved chemical substances. These chemical substances can react locally with themselves, flow between neighboring spatial regions, and can influence the behavior of mechanical processes. This formalism for point-like particles is based on the transport dissipative particle dynamics (tDPD) work by Li et. al. \cite{Li:2015fnb}. 

When we consider \emph{both} the mechanical and chemical aspects of physical or biological system, we can write the net time evolution, we observe that the \emph{state} of every object has both mechanical and chemical state variables. The mechanical state variables are of course the position, momentum, orientation and angular momentum of every discrete object. We define the mechanical state, and time evolution of each agent as:
\begin{equation}
   \frac{d^2\mathbf{r}_i}{dt^2} = \frac{d \mathbf{v}_i}{dt} = \sum_{i \neq j}
   \underbrace{
   \left( \mathbf{F}^C_{ij} + \mathbf{F}^D_{ij} + \mathbf{F}^R_{ij} \right)
   }_{\text{implicit}}
   + 
   \underbrace{\mathbf{F}^{B}_{ij}}_{\text{explicit}} + \mathbf{F}^{ext}_i,
   \label{eqn:force}
\end{equation}
where  $t$, $\mathbf{r}_i$, $\mathbf{v}_i$,
$\mathbf{F}$ are time, position velocity, and force vectors,
respectively, and $\mathbf{F}_{ext}$ is the external force on particle
$i$. Forces $\mathbf{F}^C_{ij}$, $\mathbf{F}^D_{ij}$ and
$\mathbf{F}^R_{ij}$ are the conservative, dissipative and random
forces respectively, $\mathbf{F}^{B}_{ij}$ is the \emph{bonded forces}, and  $\mathbf{F}^{ext}_i$ are the external forces. The first set of forces are what we call \emph{implicit} and the run-time automatically applies these two object pairs whose respective types and distance match the force definition. The bonded forces exist between explicit object instances, hence we refer to them as explicit forces. Finally the external force could be defined on a per-object case, in which it is explicit, or on a per-type basis, in which case we consider it implicit. In this section, we will cover model construction first using implicit force rules, and later cover defining explicit forces (bonds or links) between object instances. 

The functional form of these forces is up to the user to decide, we provide a suite of built-in force functions, and users are free to chose from, and parameterize them to best suit their modeling tasks. In this paper we will demonstrate the use of a few of these. 

As each object can represent a region of space, instead of just a simple atom, Mechanica allows users to attach a chemical cargo to each particle, and host a chemical reaction network at each element. Furthermore, we allow users to write fluxes between particles. A flux defines a movement of material from one agent to another. Furthermore, we also allow users to attach their own handlers to a variety of different events that particles (or other objects) can emit. We cover creating chemical flux networks later in detail in section, but briefly, the rate of change of the chemical concentration vector at each individual object is equal to sum of the net diffusive flux, $Q^D_{ij}$, random flux, $Q^R_{ij}$ and source term, or local reaction $Q^S_i$ as:

\begin{equation}
   \frac{dC_i}{dt} = Q_i = \sum_{i \neq j} \left (Q^D_{ij} + Q^R_{ij} \right) +
   Q^S_i,
   \label{eqn:flux}
\end{equation}
where the vector $C_i$ is the chemical state vector at each agent. The rate of change is equal to the net flux $Q_i$ which we define shortly.

\subsection{Example Applications}

\label{particles}
\subsubsection{Simple Particle Model}

We first describe the basic object type which describes a region of space. In Mechanica, we call this basic region a \emph{Particle}. A physical region of space implies that there are some local \emph{state variables} associated with it, such as location, size, density, or \emph{kind} of material. This material can be an atom, or other point particle, a molecule, fluid parcels, or a complete biological cell.

To create a particle type, we create a \emph{class} in Python based on a Mechanica particle type. For example:
\begin{minted}{python}
import mechanica as m
class MyParticleType(m.Particle): 
    pass
\end{minted}
The class {\tt MyParticleType} only describes the type, to create an object of that type we must {\em instantiate} it. This is done as follows in this case also with an optional position vector:  
\begin{minted}{python}
MyParticleType([5, 5, 5])
\end{minted}
This creates a single instance of our object, with all of it's attributes set to default values, but this isn't very useful until we attach \emph{dynamics}. An instance of a Mechanica object will not do anything, it will not change or interact with other objects until we define and attach some dynamic processes. We currently support a variety of different kinds of dynamics, that cause change in an objects position, or local state. First we start with the most basic kind of dynamics that cause objects to move. 

In order make any object in Mechanica move, we must apply a force to it. To make objects move, Mechanica sums up all of the forces that act on an object, and uses that to calculate the object’s velocity and position.

The nature of forces in Mechanica is flexible, but we provide a variety of built-in forces to enable common behaviors. The simplest kind of force we discuss is a conventional \emph{conservative} force, which we call a \emph{potential}. Long-range, fluid, and most bonded interactions are examples of conservative potential energy function based forces. All potential based forces contribute to the total potential energy of the system, and we can read the total potential energy either via the Universe.potential\_energy attribute, or we can also read the potential energy of all objects of a type, via the type’s potential\_energy attribute.

We make it easy to create forces, and apply them to objects:
\begin{minted}{python}
# create a Morse potential
p = m.Potential.morse(d=100, a=1, max=5)

# bind it to ALL instances of the MyParticleType
m.bind(p, MyParticleType, MyParticleType)
\end{minted}
This example creates a simple potential, and binds it to ALL instances of our \inpython{MyParticleType}. All objects in our modeling world are either an instance of the base Particle type, or a instance of a subclass of it. Whenever the Mechancia detects a pair of objects of this type near each other, it automatically applies this interaction between them. 

This is an example of what we call an \emph{implicit} interaction, or what's commonly called a \emph{non-bonded} interaction in molecular dynamics. We can create a complete Mechanica model, and run it using the following python script"

\begin{minted}{python}
import mechanica as m

m.init(dim=[20., 20., 20.], cutoff=3)

class B(m.Particle):
    mass = 1
    dynamics = m.Overdamped

pot  = m.Potential.morse(d=1, a=3, max=3)

m.bind(pot, B, B)

p1 = B(center + (-2, 0, 0))
p2 = B(center + (2, 0, 0))
p1.radius = 1
p2.radius = 2

m.Simulator.run()
\end{minted}
The script will open up the Mechanica interaction window when run in desktop mode, or a Jupyter notebook cell. In the desktop version, we get the following window, in Figure~\ref{fig:desktop}: 
\begin{figure}[h]
\centering
\includegraphics[width=0.6\textwidth]{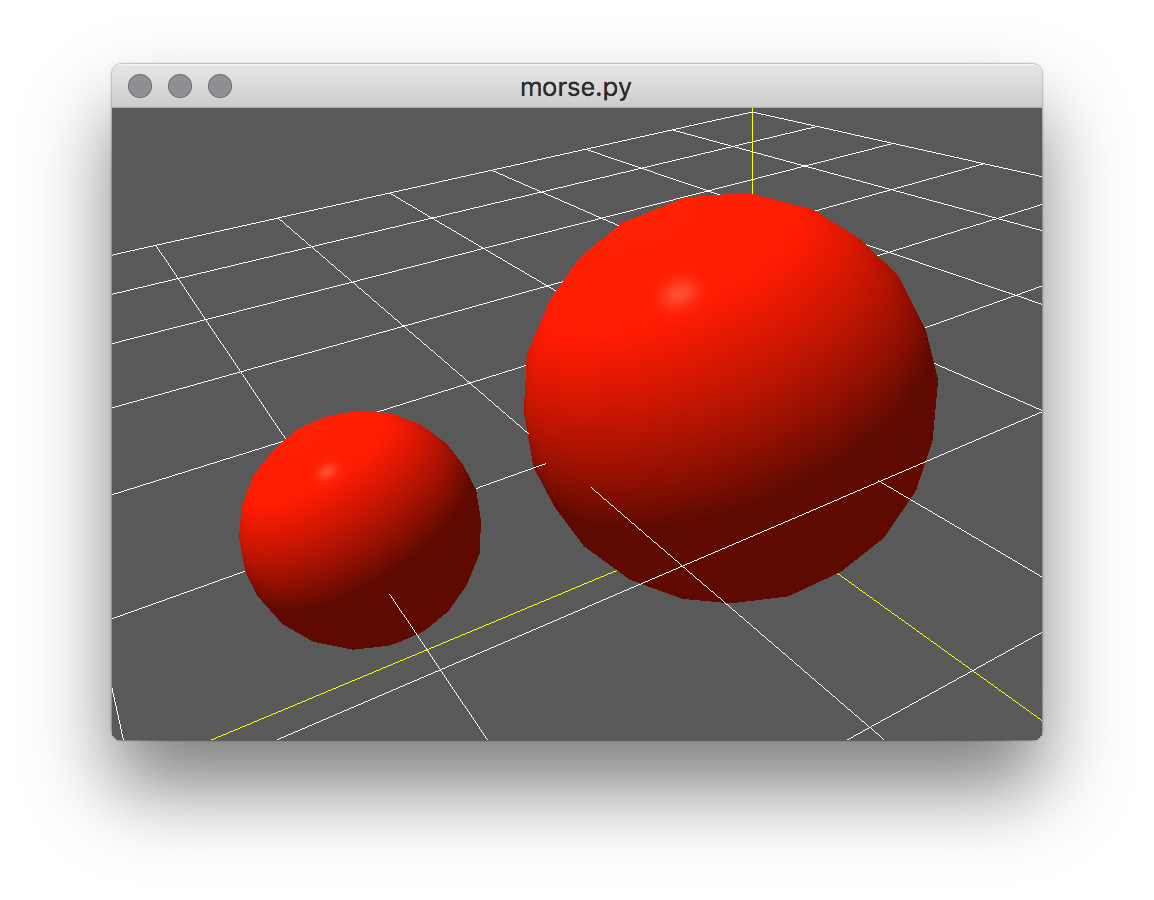}
\caption{Two particles with a Morse interaction potential, displayed in Mechanica's desktop user interface.}
\label{fig:desktop}
\end{figure}

We will notice that in the class definition of the \inpython{B} particle type, we've added a few terms in the class definition, \inpython{mass = 1}, and \inpython{dynamics = m.Overdamped}. These are examples of type level attributes, and we can have this simple syntax because we leverage the Python language \emph{Metaclass} capability. In Python, everything in an instance of a type, including the type itself. We created a special \inpython{ParticleType} class which extends the basic Python type class with a number of special features. The most directly visible feature is that all \inpython{Particle} derived classes support a range of keywords in the class definition, this is the main way users create custom classes to implement thier models. \texttt{mass} and \texttt{dynamics} are just two examples, for a complete listing of all the supported keywords, consult the Mechanica online documentation. 

The \texttt{mass} keyword here, simply defines the initial mass of each instance of the \inpython{B} type to be one, and the \inpython{dynamics} keyword instructs the Mechanica simulator that to solve the equations of motion using over-damped dynamics, i.e. velocity is equal to total applied force divided by the object's mass. We can also use standard Newtonian dynamics on any object instance simply be setting the dynamics to \texttt{Newtonian}. 

\subsubsection{Cell Sorting }

The \emph{type} of the object is the core of the Mechanica rules engine. The runtime will automatically apply a non-bonded force to a pair of objects when their type matches a given criteria, and they are close to each other. A widely studied phenomena in computation biology is cell sorting. Here we use cell sorting as a motivating example to develop a mechanica model. 

Cell sorting between biological cells of different types is one of the basic mechanisms creating tissue domains during development and wound healing and in maintaining domains in homeostasis. In this kind of model, we use the particle concept to represent biological cells as in the well-established \emph{center models}, where we treat biological cells as squishy sphere like objects.  Cells of two different types, when dissociated and randomly mixed can spontaneously sort to reform ordered tissues. Cell sorting can generate regular patterns such as checkerboards, engulfment of one cell type by another, or other more complex patterns \cite{Steinberg:1970bq}

In a classic in-vitro cell sorting experiment to determine relative cell adhesivities in embryonic tissues, mesenchymal cells of different types are dissociated, then randomly mixed and reaggregated. Their motility and differential adhesivities then lead them to rearrange to reestablish coherent homogeneous domains with the most cohesive cell type surrounded by the less-cohesive cell types \cite{Armstrong:1972ep} \cite{Armstrong:1984tc}.

Cell-sorting behavior of cell aggregates is similar to liquid surface tension, in the spontaneous separation of immiscible liquids (water vs. oil). Adhesive forces between mixed cells play a similar role in cell sorting that intermolecular attractive (cohesive) forces play in liquid surface tension. In cell sorting, the cells with the strongest highest adhesivities will be sorted to the center, while the less cohesive ones will remain outside.

To develop a computational model of our biological system, we must first identify the key objects and processes of our physical system. If we look at the left side of the following figure, we can see a sheet of biological cells. From the previous description of the cell sorting, we also know that cells move about. We know that epithelial sheets are essentially a sheet of cells that form a surface. Here we can identify our first biological object, a cell. From the previous discussion, we know that there are more than one type of cell, so lets call our two cell types, cell type A and cell type B.

First we start a model with some standard boilerplate to set up the simulation domain:

\begin{minted}{python}
import mechanica as m
import numpy as np

# total number of cells
A_count = 5000
B_count = 5000

# dimensions of universe
dim=np.array([20., 20., 20.])

# new simulator, don't load any example
m.init(dim=dim, cutoff=3)
\end{minted}
Define two different Python types to correspond to our biological cell types:
\begin{minted}{python}
class A(m.Particle):
  mass = 1
  radius = 0.5
  dynamics = m.Overdamped

class B(m.Particle):
  mass = 1
  radius = 0.5
  dynamics = m.Overdamped
\end{minted}
To represent the cell interactions, we create three different interaction potentials, \inpython{pot_aa}, \inpython{pot_bb}, and \inpython{pot_ab}. These represent the strength of interaction between cell types:
\begin{minted}{python}
# create three potentials, for each kind of particle interaction
pot_aa = m.Potential.soft_sphere(kappa=400, epsilon=40, r0=1.5, \
                               eta=2, tol = 0.05, min=0.01, max=3)

pot_bb = m.Potential.soft_sphere(kappa=400, epsilon=40, r0=1.5, \
                               eta=2, tol = 0.05, min=0.01, max=3)

pot_ab = m.Potential.soft_sphere(kappa=400, epsilon=5, r0=1.5, \
                               eta=2, tol = 0.05, min=0.01, max=3)
\end{minted}
And bind those potentials to our cell types:
\begin{minted}{python}
# bind the potential with the *TYPES* of the particles
m.bind(pot_aa, A, A)
m.bind(pot_bb, B, B)
m.bind(pot_ab, A, B)
\end{minted}
In over-damped dynamics, we need a random force to enable the objects to move around, otherwise they tend to get trapped in a potential wells. This random force represents basically a cell motility:
\begin{minted}{python}
rforce = m.forces.random(0, 50)

# bind it just like any other force
m.bind(rforce, A)
m.bind(rforce, B)
\end{minted}
Create the particle instances:
\begin{minted}{python}
# create particle instances, for a total A_count + B_count cells
for p in np.random.random((A_count,3)) * 15 + 2.5:
  A(p)
for p in np.random.random((B_count,3)) * 15 + 2.5:
  B(p)
# finally run the simulation:
m.Simulator.run()
\end{minted}
The output of this simulation is shown in Figure~\ref{fig:cell-sorting} below. 
\begin{figure}[h!]
\centering
\includegraphics[width=0.5\textwidth]{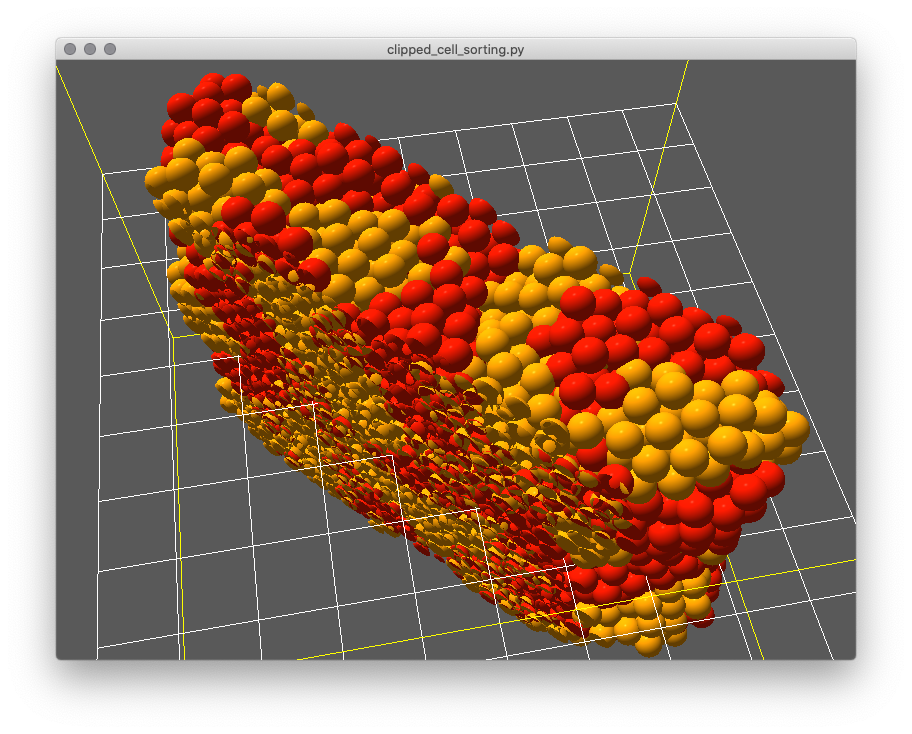}
\caption{A simple cell sorting simulation demonstrating Mechanica used as a \emph{center model} simulator. This figure shows the use of user defined clipping planes, we support up to eight clipping planes of arbitrary orientation. Users can change the orientation of the clipping planes at runtime, and we automatically copy these orientation over the GPU, where a pixel shader clips objects outside the cutting planes. We are currently working on improving the visual appearance of the clipping planes where we will cap the surfaces of any clipped objects. }
\label{fig:cell-sorting}
\end{figure}

\subsubsection{Fluids: Poiseuille Flow}

Fluids in Mechanica are another example of implicit interactions, but the interaction force here adds an additional component to enable momentum diffusion between objects, to represent dissipation. In fact, Mechanica's fluid model is an implementation of the well-established dissipate particle dynamics (DPD) method. In this kind of model, we use particles to represent small `parcels` of fluid. Briefly, the DPD force is the sum of the conservative, dissipative and random forces:
\begin{align}
    \mathbf{F}_{ij} &= \mathbf{F}^C_{ij} + \mathbf{F}^D_{ij} + \mathbf{F}^R_{ij}, \\
    \mathbf{F}^C_{ij} &= \alpha \left(1 - \frac{r_{ij}}{r_c}\right)\mathbf{e}_{ij} \\
    \mathbf{F}^D_{ij} &= -\gamma \left(1 - \frac{r_{ij}}{r_c}\right)^{2}(\mathbf{e}_{ij} 
        \cdot \mathbf{v}_{ij}) \mathbf{e}_{ij}\\
    \mathbf{F}^R_{ij} &= \sigma \left(1 - \frac{r_{ij}}{r_c}\right) 
        \xi_{ij}\Delta t^{-1/2}\mathbf{e}_{ij},
\end{align}
where the conservative force $\mathbf{F}^C_{ij}$ represents the inertial forces in the fluid, the dissipative, or friction force $\mathbf{F}^D$ represents the tendency for fluid parcel momentums to align, and the random force $\mathbf{F}^R$ is a pair-wise random force between particles. Users are of course free to choose any forces they like, but these are the most commonly used DPD ones. 

Here we write a model of Poiseuille flow, which is a pressure-driven flow (Channel Flow) in a tube or pipe. We start a fluid model just like the others, first with some boiler plate, but now we introduce Mechanica's flexibly boundary conditions. We support a range of standard kind of boundary conditions and these are easily specified using the \inpython{bc} keyword argument to the init function. Please refer to the docs for a complete list of supported boundary conditions, but this example uses periodic in the X and Y directions, and sets the top and bottom planes to zero velocity. We can also attach potentials to boundary planes to make them repulsive or attractive, and we can even attach a DPD potential to them, to help align particle momentum near the boundary in softer way than just with recoil effects. 

\begin{minted}{python}
import mechanica as m
import numpy as np

m.init(dt=0.1, dim=[15, 12, 10], cells=[7, 6, 5], cutoff=0.5,
       bc={'x':'periodic',
           'y':'periodic',
           'top':{'velocity':[0, 0, 0]},
           'bottom':{'velocity':[0, 0, 0]}})
\end{minted}
Now we create our basic particle types and set some style attributes, and create an instance of the DPD potential, and bind that potential between pairs of our particle type. 
\begin{minted}{python}
class A (m.Particle):
    radius = 0.05
    style={"color":"seagreen"}
    dynamics = m.Newtonian
    mass=10

dpd = m.Potential.dpd(alpha=10, sigma=1)
m.bind(dpd, A, A)
\end{minted}
Poiseuille flow requires a pressure to drive the flow, we represent that with a constant force that we attach to our particle types: 
\begin{minted}{python}
pressure = m.forces.ConstantForce([0.1, 0, 0]) # driving pressure
m.bind(pressure, A)
\end{minted}
Now we want to initialize our model, we use the built-in lattice initializer to create a simple cubic lattice of particle type A at the center of our simulation domain.  And finally run the simulation. 
\begin{minted}{python}
uc = m.lattice.sc(0.15, A) # 0.15. lattice spacing
parts = m.lattice.create_lattice(uc, [40, 40, 40])
m.run()
\end{minted}
The entire running simulation is shown below in Figure~\ref{fig:poiseuille}. Here we ran the model inside of Mechanica's Jupyter web front end. The Jupyter notebook front end provides a convenient user interface for interactive model construction all within a web browser. Mechanica can be run either locally, or remotely, and may be accessed via the publicly unavailable nanoHub install. The Jupyter user interface is currently not a fast as the native desktop version, but we are working on number of performance improvements which we expect to achieve interactive use.
\begin{figure}[h!]
\centering
\begin{minipage}{.48\textwidth}
  \centering
  \includegraphics[width=\linewidth]{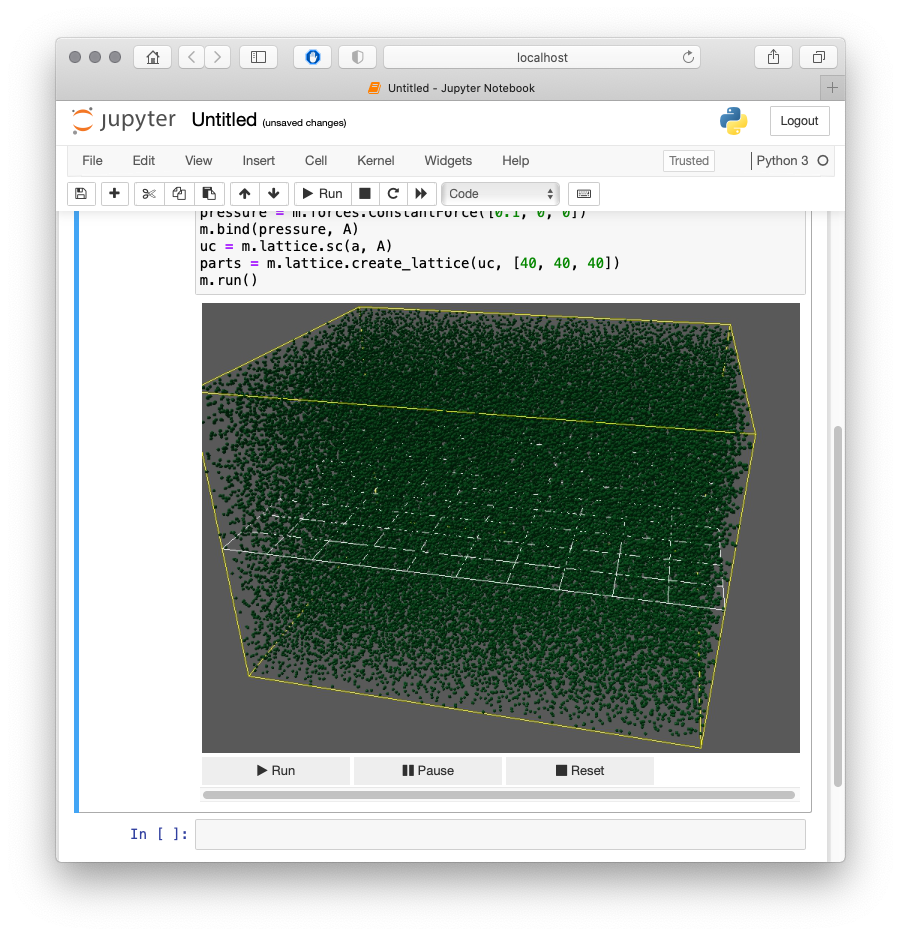}
  \label{fig:test1}
\end{minipage}%
\begin{minipage}{.48\textwidth}
  \centering
  \includegraphics[width=\linewidth]{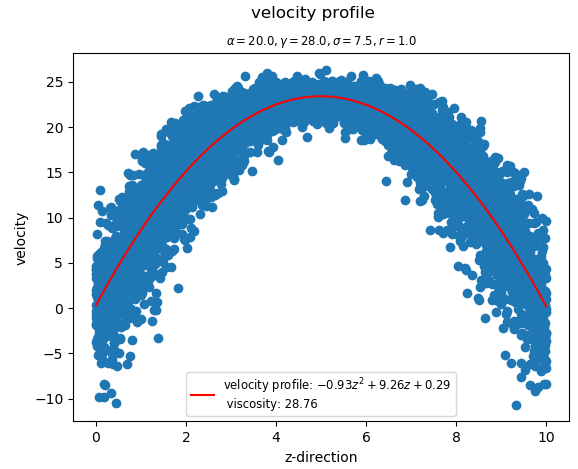}
  \label{fig:test2}
\end{minipage}
\caption{A Poiseuille flow model running a web-browser using Mechanica's Jupyter notebook front end. The simulation is live and interactive, and user is free to continue interacting with the model via altering the model structure with standard Python expressions. The right pane displays a scatter plot of the particle velocities in the X-Y plane, and a quadratic polynomial fit using Matplotlib. }
\label{fig:poiseuille}
\end{figure}

\FloatBarrier
\subsubsection{Space limited cell growth}
Biological cell growth is often space limited -- cells can sense the pressure exerted by it's environment, and when there is sufficient space (i.e. relative pressure is low), and assuming there is sufficient nutrients, cells are known to grow and expand. Cells are also known to divide divide when then they reach a certain size. We can implement these observed behaviors in a very simple Mechanica model. Lets begin with taking a look at Listing~\ref{lst:pressure}:
\begin{listing}[H]
\begin{minted}[linenos]{python}
import mechanica as m
import numpy as np

m.init(dim=[20., 20., 20.], cutoff=7)

pot = m.Potential.morse(d=0.5, a=5, max=4)

class Cell(m.Particle):
    mass = 10
    target_temperature = 0
    radius=0.2
    dynamics=m.Overdamped
    style = { "colormap" : {
        "attr" : "number_density", "map" : "rainbow","range" : "auto"}}

    def on_time(self, event):
        if self.number_density < 0.1
            self.radius += 0.1

        if self.radius > 2:
            self.split()

m.on_time(Cell.on_time, period=0.2, distribution='exponential')

m.Universe.bind(pot, Cell, Cell)

rforce = m.forces.random(0, .01)

m.Universe.bind(rforce, Cell)

Cell([10., 10., 10.])

m.Simulator.run()
\end{minted}
\caption{A Mechanica model for pressure limited cell growth}
\label{lst:pressure}
\end{listing}
This model starts out with the usual few lines of boiler plate, and we create an instance of the widely used Morse potential to implement a somewhat soft interaction between the constituent cells in lines 1-4. In line 8 we define our cell type, and set some type values. Note here that we set the \texttt{style} attribute. Mechanica styles are a general way to define the visual appearance of objects, and can be set on a type (class) or individual level. Here we tell the renderer that we want to use a color map to map a scalar value to a displayed color, and we will use the \texttt{number\_density} attribute to determine color. 

The \texttt{number\_density} $\rho(\mathbf{x}_i)$ attribute on every object is an estimate of the net \emph{pressure} at that area of space. It gets automatically computed at each time step, as the smoothing kernel weighted sum of all other objects: 
\begin{equation}
    \rho(\mathbf{x}_i) = \sum_j W(\left| \mathbf{x}_i - \mathbf{x}_j \right| / h).
\end{equation}
The choice of smoothing kernel is somewhat arbitrary, but we use the widely used cubic spline kernel of Monaghan~\cite{Monaghan:2005fd}: 

\begin{equation}
    W(q) = \begin{cases}
                \sigma_3\left[ 1 - \frac{3}{2}q^2\left( 1 - \frac{q}{2} \right) \right], 
                    & 0 \leq q \leq 1, \\
                \frac{\sigma_3}{4}(2-q)^3, & 1 < q \leq 2, \\
                 0 & q>2, \\
            \end{cases}
\end{equation}
where $q$ is the inter particle distance re-scaled to $[0,1]$, i.e. $q = x / \mathrm{cutoff}$, and  $\sigma_3 = {1}/{\pi h^3}$,  is a dimensional normalizing factor for the cubic spline function.

To implement a simple pressure limited cell growth behavior, we can add a method to the class definition, and tell the run time to call this periodically as wish the \texttt{on\_time} method on line 23. The choice of method name is arbitrary, and users can add as many time event listeners as they wish. In this method, we increase the object instance size if the number density is lower than some threshold in line 18. And if the size of the object is sufficiently large, we call the \texttt{split()} method on the object instance. The \texttt{split()} method will perform a cell mitosis, and split the object in half, dividing the contents of the parent object uniformly between the two daughter cells, and these two new cells will get automatically added to the simulation at the current location. 

\begin{figure}[h]
    \centering
    \includegraphics[width=0.5\textwidth]{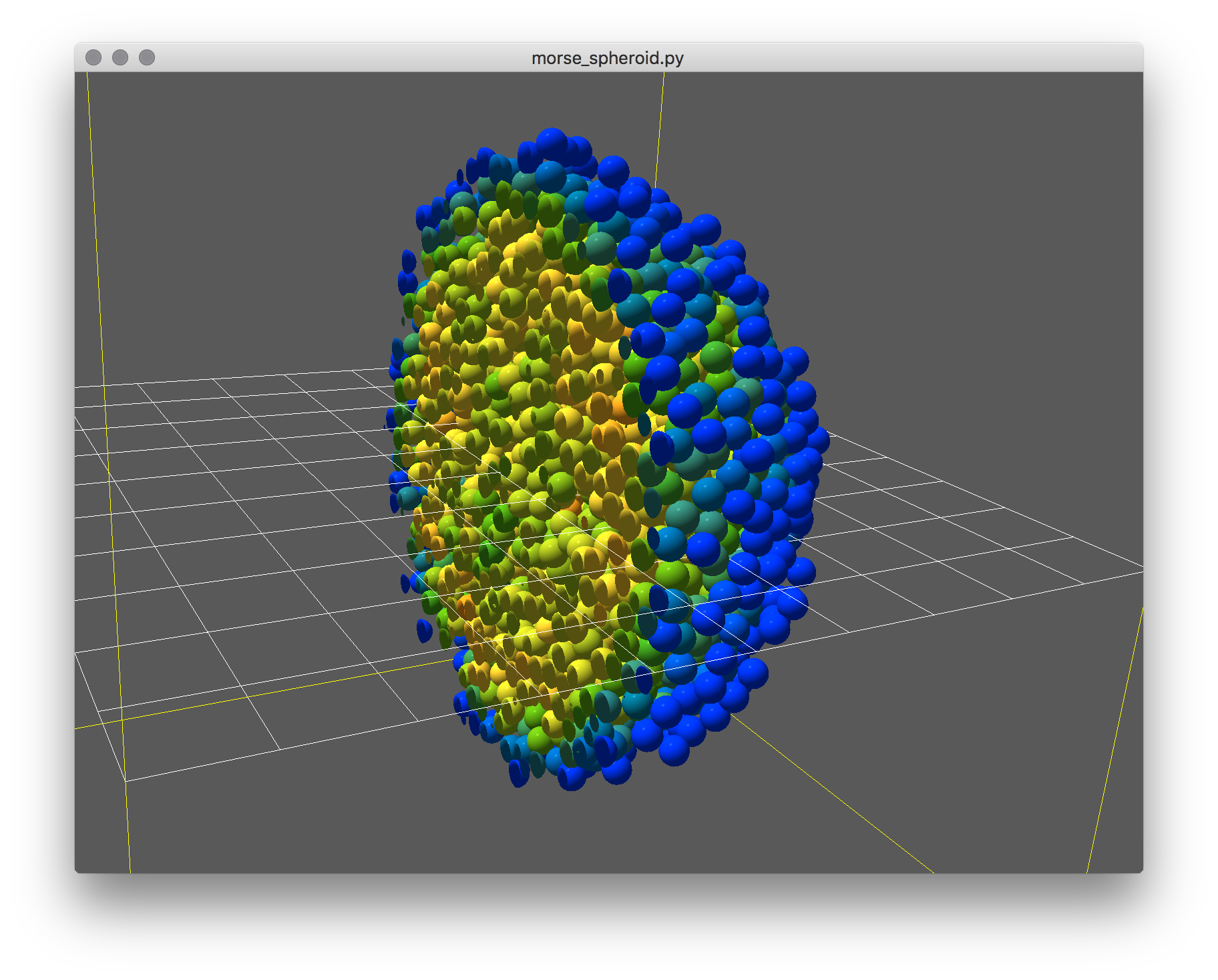}
    \caption{A simple space limited cell growth model after several thousand time steps. The outer layer has a lower density, as it only has objects on one side, thus the growth rule preferentially splits cells in this region.}
    \label{fig:morse_spheroid}
\end{figure}

\subsubsection{Clusters: Hierarchical Organization of Space}
Until now we have focused on Mechanic's most primitive object type for representing space: the point like particle. Particle's in Mechanica can also have a well-defined complex shape, as we also support discrete element model based particles, fully meshed objects, and are developing tetrahedral meshed volume elements, but for the purposes of this paper, we will only cover the next level up from point-like particles: the cluster based particle. Cluster based modeling frameworks are frequently called \emph{sub-cellular element} models, \cite{Fletcher:ca, Sweet:kl, Tarama:2019wu, Jamali:2010ww, Sandersius:2011ih}

One of our main goals in developing Mechanica is a consistent and simple model construction API across different modeling formalisms. As such, we use the exact same formalism developed earlier to bind forces and processes to clusters. The Mechanica \inpython{Cluster} class is a sub-class of the base \inpython{Particle} type, and responds to the same methods, and has the same programmatic interface.

To develop a Cluster based model, we derive our agent type definitions from the \inpython{Cluster} instead of the \inpython{Particle} base type. To define object types that will exist \emph{inside} the new cluster, we simply create a nested class, using standard Python syntax. For example, to create a cluster called \texttt{C}, which contains two different particle types that will live inside an instance of this cluster called \texttt{A} and \texttt{B}, we would:
\begin{minted}{python}
class C(m.Cluster):
    radius=3

    class A(m.Particle):
        radius=0.5
        dynamics = m.Overdamped
        mass=10
        style={"color":"MediumSeaGreen"}

    class B(m.Particle):
        radius=0.5
        dynamics = m.Overdamped
        mass=10
        style={"color":"skyblue"}
\end{minted}
This syntax is possible because we have implemented the \texttt{Particle} base type as a \emph{meta-class} in the core C++ library, as as such, we have significant control over the declaration and construction of any derived types. In this example, we have simply created two \texttt{Particle} derived types inside this \texttt{Cluster} definition. To create an instance of one of these objects inside the parent cluster, we first need to create an instance of the parent cluster type, then simply call the name of one of the nested types as:
\begin{minted}{python}
c = C() # creates an instance of the C cluster type. 
c.A()   # creates an instance of the C.A particle inside the c object
c.B()   # creates an instance of the C.B particle inside the c object. 
\end{minted}
This nested type definitions can be continued arbitrarily, we can create clusters inside of clusters. 

\begin{figure}[!h]
    \centering
    \includegraphics[width=0.5\textwidth]{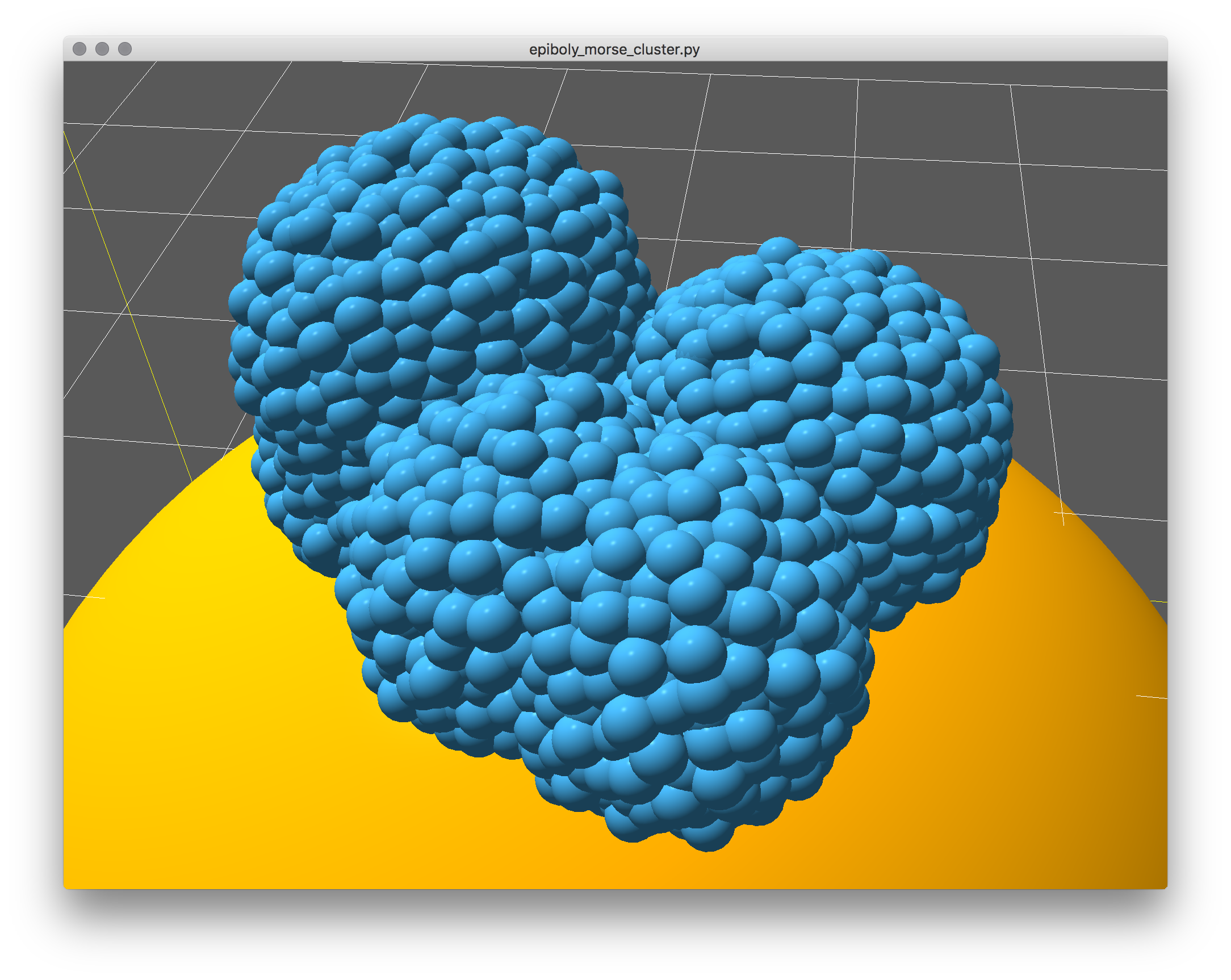}
    \caption{A \emph{cluster} based model of cell mitosis. These five cells started out as single initial cell on a "yolk" surface, and we triggered four \texttt{split()} operations. }
    \label{fig:epiboly_morse_cluster}
\end{figure}

Contained objects are subtly different that free objects in how we define their force relationships. Say we have two biological cells of the same biological cell type that are in contact with each other, and we wish to create a detained model of their internal structure, and we choose to create a \texttt{Cluster} derived type to represent our biological cell, and a \texttt{Particle} derived nested type to represent the cytoplasm. We know that the cytoplasm in one cell will interact differently with cytoplasm in the same cell, as with cytoplasm in another, in effect, they have different identities. In this intermediate level of detail model, we choose not to represent an explicit cell membrane. To make the distinction between object interactions within the same container, and interactions of the same type, but in different containers, we introduce the \emph{bound} concept. We refer to interactions of the a type within a container as bound interactions, and interactions of a type within separate containers as free. Thus, we can attach forces to these object just like we attach forces to any other object, by using it's symbol name, but we inform the run time rules engine that these are bound vs. free interactions with the \emph{bound} argument as: 
\begin{minted}{python}
p1  = m.Potential.morse(d=5, a=5, max=3)
p2  = m.Potential.morse(d=0.1, a=5, max=3)
m.bind(p1, C.A, C.A, bound=True)
m.bind(p2, C.A, C.A, bound=True)
\end{minted}
Here we create two forces, \texttt{p1} and \texttt{p2}, and we see that \texttt{p1} is stronger than \texttt{p2}. We attach these forces to the nested \texttt{C.A} type. This example results in contained objects adhering strongly within the same cluster, and less strongly with different clusters. Thus, they tend to form blob-like structures, as shown in Fig.~\ref{fig:epiboly_morse_cluster}. 
The \texttt{Cluster.split()} method has the same signature as the base class \texttt{Particle.split()}, but accepts additional arguments, and enables additional control over splitting clusters. The \texttt{split} method splits a cluster into itself, and a new daughter cluster. Split accepts optional normal and point arguments to define a cleavage plane. As clusters are spatially extended objects, with complex geometry, we need more control. The \texttt{Cluster.split} method can accept an optional a point and normal vector to define a cleavage plane. In this case, we enumerate all of the constituent objects of the cluster, and separate them based on which side of the cleavage plane they are and split them. For example, to create a cell instance, and split it along a normal vector (if we only specify a normal vector, the split method uses the center of mass of the cluster as the point): 
\begin{minted}{python}
c = MyCell(...)
d = c.split(normal=[1., 0., 0.])
\end{minted}
where d is the new daughter cell. This form uses the center of mass of the cluster as the cleavage plane point. We can connect the split method to events just like in our space limited growth model as shown in Fig.~\ref{fig:epiboly_morse_cluster}.

\subsubsection{Spatial Transport and Flux}
The concept of a flux is extremely general, and this lets us define a connector type that lets users connect different model elements. Flux networks allow us to define a wide range of problems, from biological fluid flow in areas like the liver and the eye, to physiologically based pharmacokinetic (PBPK) modeling, and even to electric circuits and pipe flow networks. 

Unlike a traditional micro-scale molecular dynamics approach, where each
computational particle represents an individual physical atom, a DPD is
mesoscopic approach, where each computational particle represents a 'parcel' of
a real fluid. A single DPD particle typically represents anywhere from a cubic
micron to a cubic mm, or about $3.3 \times 10^{10}$ to $3.3 \times
10^{19}$ water molecules.

Transport dissipative particle dynamics (tDPD) adds diffusing chemical solutes
to each classical DPD particle. Thus, each tDPD particle represents a parcel of
bulk fluid (solvent) with a set of chemical solutes at each particle. In tDPD,
the main particles represent the bulk medium, or the 'solvent', and these carry
along, or advect attached solutes. We introduce the term 'cargo' to refer to the
localized chemical solutes at each particle.

In general, the time evolution of the chemical species at each spatial object 
is given as:

\begin{equation}
   \frac{dS_i}{dt} = Q_i = \sum_{i \neq j} Q^T_{ij} +Q^R_i,
\end{equation}
where the rate of change of the vector of chemical species at an object is equal
to the flux vector, $Q_i$. This is the sum of the transport and reactive
fluxes. $Q^T$, is the \emph{transport flux}, and $Q^R_i$ is a local
reactive flux, in the form of a local reaction network. We will cover local
reaction fluxes in later paper, for now we restrict this discussion to the passive or 'Fickian' flux, \emph{secretion} flux and \emph{uptake} flux.

Before we cover spatial transport, we first cover adding chemical reaction
networks to objects. To attach chemical cargo to a particle, we simply add a ``species`` specifier to
the particle type definition as
\begin{minted}{python}
class A(m.Particle):
    species = ['S1', 'S2', S3']
\end{minted}
This defines the three chemical species, `S1`, `S2`, and `S3` in the *type*
definition. Thus, when we create an \emph{instance} of the object, that instance will
have a vector of chemical species attached to it, and is accessible via the
\texttt{Particle.species} attribute. Internally, we allocate a memory block for
each object instance, and users can attach a set of reactions to define the time
evolution of these attached chemical species.
If we access this species list from the \emph{type}, we get:
\begin{minted}{python}
>>> print(A.species)
SpeciesList(['S1', 'S2', 'S3'])
\end{minted}
This is a special list of SBML species definitions. It's important to note that
once we've defined the list of species in each time, that list is
immutable. Creating a list of species with just their names is the simplest
example, if we need more control, we can create a list from more complex species
definition strings in \texttt{Species}.

If a type is defined with a species definition, every instance of that
type will get a StateVector, of these substances. Internally, a state vector
is really just a contiguous block of numbers, and we can attach a reaction
network or rate rules to define their time evolution. 

Each instance of a type with a `species` identifier gets a `species`
attribute, and we can access the values here. In the instance, the `species`
attribute acts a bit like an array, in that we can get it's length, and use
numeric indices to read values:
\begin{minted}{python}
>>> a = A()
>>> print(a.species)
StateVector([S1:0, S2:0, S3:0])
\end{minted}
As we can see, the state vector is array like, but in addition to the numerical
values of the species, it contains a lot of meta-data of the species
definitions. We can access individual values using array indexing as:
\begin{minted}{python}
>>> print(a.species[0])
0.0

>>> a.species[0] = 1
>>> print(a.species[0])
1.0
\end{minted}
The state vector also automatically gets accessors for each of the species
names, and we can access them just like standard Python attributes:
\begin{minted}{python}
>>> print(a.species.S1)
1.0

>>> a.species.S1 = 5
>>> print(a.species.S1)
5.0
\end{minted}
We can even get all of the original species attributes directly from the
instance state vector like:
\begin{minted}{python}
>>> print(a.species[1].id)
'S2'
>>> print(a.species.S2.id)
'S2'
\end{minted}
In most cases, when we access the species values, we are accessing the
*concentration* value. See the SBML documentation, but the basic idea is that we
internally work with amounts, but each of these species exists in a physical
region of space (remember, a particle defines a region of space), so the value
we return in the amount divided by the volume of the object that the species is
in. Sometimes we want to work with amounts, or we explicitly want to work with
concentrations. As such, we can access these with the `amount` or `conc`
attributes on the state vector objects as such::

\begin{minted}{python}
>>> print(a.species.amount)
1.0

>>> print(a.species.conc)
0.5
\end{minted}

This simple version of the \texttt{species} definition defaults to create a set of
\emph{floating} species, or species who's value varies in time, and they participate
in reaction and flux processes. We also allow other kinds species such as
\emph{boundary}, or have initial values. 

The Mechanica \texttt{Species} object is essentially a Python binding around the
libSBML Species class, but provides some Pythonic conveniences. For example, in
our binding, we use conventional Python `snake\_case` sytax, and all of the sbml
properties are available via simple properties on the objects. Many SBML
concepts such as `initial\_amount`, `constant`, etc. are optional values that may
or may not be set. In the standard libSBML binding, users need to use a variety
of `isBoundaryConditionSet()`, `unsetBoundaryCondition()`, etc... methods that
are a direct python binding to the native C++ API. As a convience to our
users, our methods simply return a Python `None` if the field is not set,
otherwise returns the value, i.e. to get an initial amount:

\begin{minted}{python}
>>> print(a.initial_amount)
None
>>> a.initial_amount = 5.0
\end{minted}

This internally updates the libSBML `Species` object that we use. As such, if
the user wants to save this sbml, all of these values are set accordingly. 

The simplest species object simply takes the name of the species as the only
argument:

\begin{minted}{python}
>>> s = Species("S1")
\end{minted}

We can make a \emph{boundary} species, that is, one that acts like a boundary
condition with a "\$" in the argument as:

\begin{minted}{python}
>>> bs = Species("$S1")
>>> print(bs.id)
'S1'
>>> print(bs.boundary)
True
\end{minted}

The Species constructor also supports initial values, we specify these by adding
a \texttt{= value} right hand side expression to the species string:

\begin{minted}{python}
>>> ia = Species("S1 = 1.2345")
>>> print(ia.id)
'S1'
>>> print(ia.initial_amount)
1.2345
\end{minted}

\subsubsection{Advection}
Recall that the bulk or solvent particles don't represent a single molecule,
but rather a parcel of fluid. As such, dissolved chemical solutes (cargo) in each
parcel of fluid have natural tendency to *diffuse* to nearby locations.
\begin{figure}[h]
    \centering
    \includegraphics[width=0.5\textwidth]{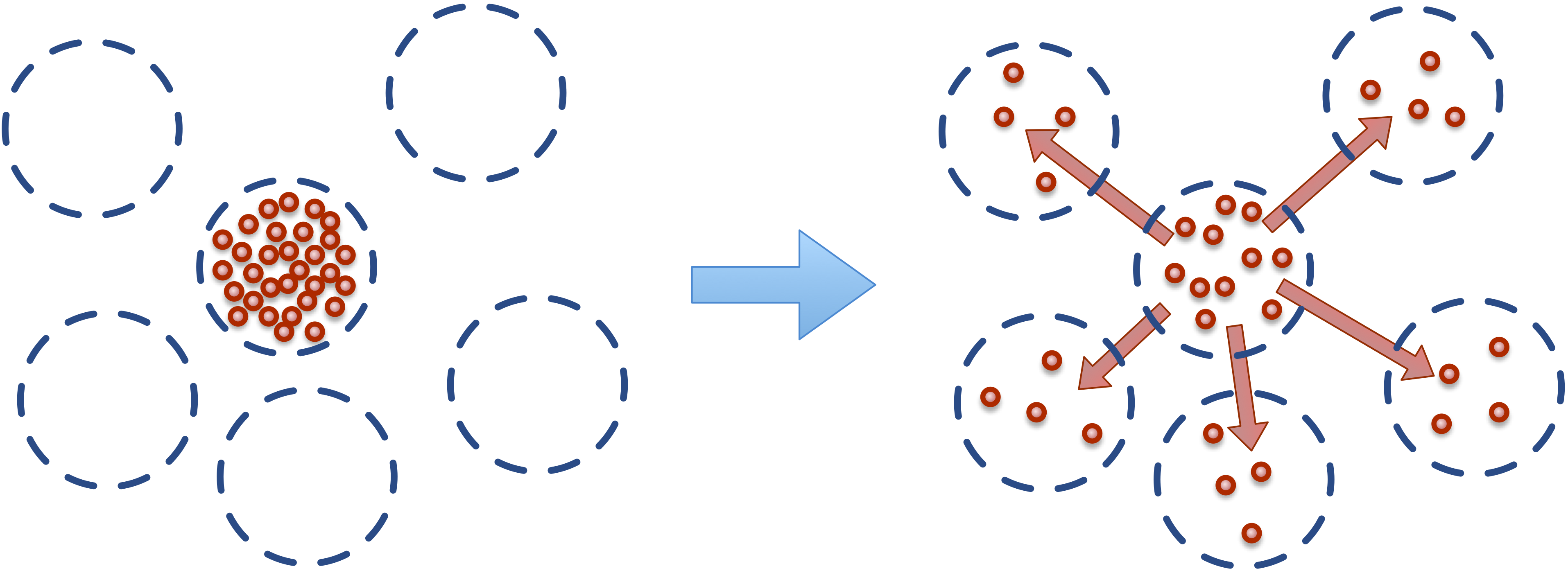}
    \caption{Dissolved solutes have a natural tendency to diffuse to nearby locations.}
    \label{fig:diffusion}
\end{figure}

This micro-scale diffusion of solutes results in mixing or mass transport
without directed bulk motion of the solvent. We refer to the bulk motion, or
bulk flow of the solvent as \emph{advection}, and use \emph{convection} to describe the
combination of both transport phenomena. Diffusion processes are typically
either normal or anomalous. Normal (Fickian) diffusion obeys Fick's laws,
and anomalous (non-Fickian) does not.

We introduce the concept of \emph{flux} to describe this transport of material
(chemical solutes) between particles. Fluxes are similar
similar to conventional pair-wise forces between particles, in that a flux is
between all particles that match a specific type and are within a certain
distance from each other. The only differences between a flux and a force, is
that a flux is between the chemical cargo on particles, and modifies
(transports) chemical cargo between particles, whereas a force modifies the net
force acting on each particle.

We attach a flux between chemical cargo as::

\begin{minted}{python}
class A(m.Particle)
 species = ['S1', 'S2', 'S3']

class B(m.Particle)
 species = ['S1, 'Foo', 'Bar']

q = m.fluxes.fickian(k = 0.5)

m.bind(q, A.S1, B.S)
m.bind(q, A.S2, B.Bar)
\end{minted}

This creates a Fickian diffusive flux object \texttt{q}, and binds it between species
on two different particle types. Thus, whenever any pair of particles instances
belonging to these types are near each other, the run-time will apply a Fickian
diffusive flux between the species attached to these two particle instances. 

\subsubsection{Passive Flux: Diffusion}

We implement a diffusion process of chemical species located at object instances
using the basic passive (Fickian) flux type, with the \texttt{flux}. This flux
implements a passive  transport between a species located on a pair of nearby objects of type a
and b. A Fick flux of the species \texttt{S} attached to object types
\texttt{A} and \texttt{B} implements the reaction:

\begin{eqnarray}
a.S & \leftrightarrow a.S \; &; \; k \left(1 - \frac{r}{r_{cutoff}} \right)\left(a.S - b.S\right)     \\
a.S & \rightarrow 0   \; &; \; \frac{d}{2} a.S \\
b.S & \rightarrow 0   \; &; \; \frac{d}{2} b.S,
\end{eqnarray}

$S$ is a chemical species located at each
object instances. :math:`k` is the flux constant, :math:`r` is the
distance between the two objects, $r_{cutoff}$ is the global cutoff
distance, and $d$ is the optional decay term.

\subsubsection{Active Fluxes: Production and Consumption}

For active pumping, to implement such processes like membrane ion pumps, or
other forms of active transport, we provide the \texttt{produce\_flux} and
\texttt{consume\_flux} objects. The produce flux implements the reaction:
\begin{eqnarray}
   a.S & \rightarrow b.S \; &; \;  k \left(r - \frac{r}{r_{cutoff}} \right)\left(a.S - a.S_{target} \right) \\
   a.S & \rightarrow 0   \; &; \;  \frac{d}{2} a.S \\
   b.S & \rightarrow 0   \; &; \;  \frac{d}{2} b.S,
\end{eqnarray}
and the consumer flux implements the reaction:
\begin{eqnarray}
a.S & \rightarrow b.S \; &; \; k \left(1 - \frac{r}{r_{cutoff}}\right)\left(b.S - b.S_{target} \right)\left(a.S\right) \\
a.S & \rightarrow 0   \; &; \; \frac{d}{2} a.S \\
b.S & \rightarrow 0   \; &; \; \frac{d}{2} b.S
\end{eqnarray}
Here, the $\left(1 - \frac{b.S}{b.S_{target}} \right)$ influences the
forward rate, where $[b.S]$ is the concentration of the substance S, and
$b.S_{target}$ is the target concentration. The flux will continue forward
so long as there is both concentration of the reactant, $a.S$, and
the product $b.S$ remains below its target value. Notice that if the
present concentration of $b.S$ is \emph{above} its target, the reaction will
proceed in the reverse direction. Thus, the \texttt{pumping\_flux} can be used to
implement both secretion and uptake reactions.

\subsubsection{Explicit Bonded Connections}
We have up till now only covered implicit interactions, again, implicit interactions are transient and are active when objects of a specific type have a specific spatial relationship. These are great for modeling substances such as fluids, gasses, but when we look at more complex materials such as molecules, visco-elastics or biological materials, we need semi-persistent explicit bonded relationships.  Mechanica supports the standard suite of bonds commonly used in molecular dynamics, and we add certain attributes such as breaking energy, so that bonds can break if stretched too far. We create a explicit bond in Mechanica just like we create implicit relationships, using the \texttt{bind} function, but instead of object types as the second and third augments, we instead pass in a pair of object instances.

We create a model of a visco-elastic solid similarly to how we created earlier fluid models, but use some different interaction potentials, and a new explicit bond between neighboring particle. We create this bonded mesh between the objects in their initial configuration, so that in response to small perturbations, the material will always return to it's initial, or rest configuration. If however the bond stretches too far, it will tear, and this the reference configuration of the material will change. Here we create a model with three kids of materials, \texttt{A, B, Fixed}. \texttt{Fixed} is a boundary type material which does not move, and only serves to anchor the main material to. \texttt{B} is the leading edge material that we attach sinusoidal driving force to, and \texttt{A} is the main material. In this model, we create a cubic blob of material, arranged in a simple cubic lattice, and explicit bonds along the lattice basis vectors. Then we attach the left side of the material to a rigid wall, and apply a sinusoidal driving force to the right side.  The driving force is periodic parallel to the rigid wall, and pulls uniformly perpendicular to the wall. Thus the force wants to pull directly on the material, while rocking it back and forth, this would be similar if way one wants to dislodge a piece of sticky material from a surface. We create this model just like previous ones with first creating particle types to represent our respective material types.

\begin{figure}[h]
\centering
\includegraphics[width=0.6\textwidth]{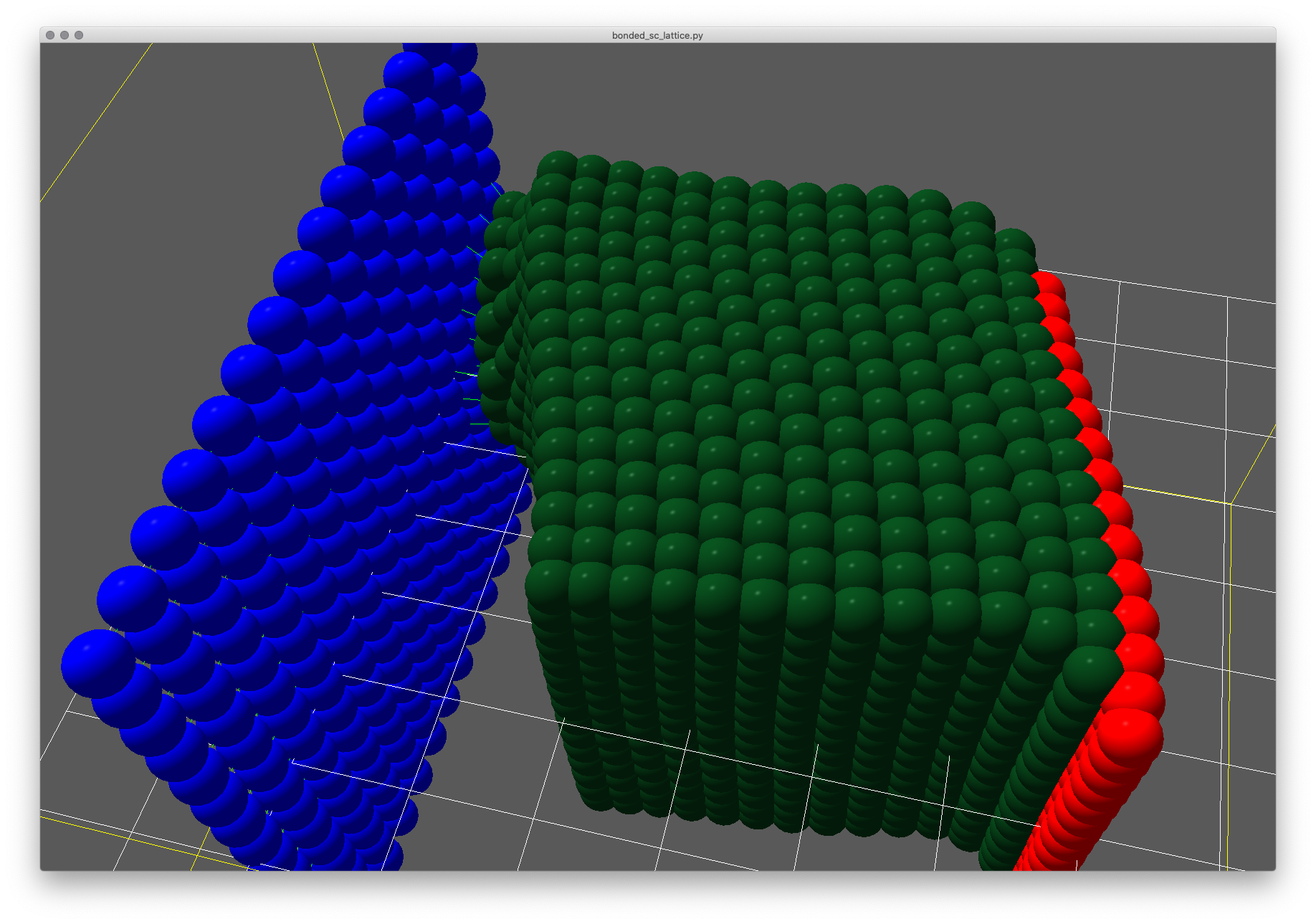}
\caption{Screenshot of a visco-elastic material simulation}
\label{fig:visco-elastic}
\end{figure}

\begin{minted}{python}
import mechanica as m
import numpy as np

m.init(dt=0.1, dim=[15, 12, 10]) 

class A (m.Particle):
    radius = 0.3
    style={"color":"seagreen"}
    dynamics = m.Overdamped

class B (m.Particle):
    radius = 0.3
    style={"color":"red"}
    dynamics = m.Overdamped

class Fixed (m.Particle):
    radius = 0.3
    style={"color":"blue"}
    frozen = True
\end{minted}
We attach a repulsive Coulomb force in the material, this essentially provides an internal pressure to the material, while the bonded forces tend to keep the material together. We create a sinusoidal driving force, and attach that to the leading edge material. Note here, we use a Python \inpython{lambda} expression to define this function. What the Mechanica runtime does is re-sample this Python defined expression at regular intervals, save that value and apply it to the simulation objects, as it would be far too costly to evaluate the Python expression for each object. We are presently working on JIT compiling certain subsets of Python which to use for user-defined forces, that will be incorporated in a future version. 
\begin{minted}{python}
repulse = m.Potential.coulomb(q=0.08, min=0.01, max=2*a)
m.bind(repulse, A, A)
m.bind(repulse, A, B)
f = m.forces.ConstantForce(lambda: [0.3, 1 * np.sin( 0.4 * m.Universe.time), 0], 0.01)
m.bind(f, B)
\end{minted}
Now we create the explicit bonded interactions. We create an instance of the generalized power law force to use as the potential function for the bonds. Here we use the lattice initializer, and pass it a function which yields a bond that the initializer than connects respective object instance pairs. We specify an optional dissociation energy for the bonds, this is the maximum strength of the bond, and the runtime looks at the bond energy at each step, and if the energy exceeds this specified threshold, the runtime automatically removes the bond. We initially created the material as a uniform solid with every element of type \texttt{A}, in the last two lines, we iterate over all the left and right planes of the lattice and change their type to either \texttt{Fixed} or \texttt{B} respectively. Users are free to dynamically change the type of an object instance at any time during the simulation. 
\begin{minted}{python}
pot = m.Potential.power(r0=0.5*a, alpha=2)
uc = m.lattice.sc(
    0.65, A, lambda i, j: m.Bond(pot, i, j, dissociation_energy=1.3))
parts = m.lattice.create_lattice(uc, [15, 15, 15])
for p in parts[0,:].flatten(): p[0].become(Fixed)
for p in parts[14,:].flatten(): p[0].become(B)
m.run()
\end{minted}
And finally run the model, who's output is show in Figure~\ref{fig:visco-elastic}. 

\subsection{Simulator Architecture and Multi-Platform Rendering}
\label{sec:arch}

Here we will briefly discuss the internal simulator and rendering architecture.  

We partition all space into a regular set of spatial \emph{voxels}, usually around 5x5x5 total, and agents can move between voxels. Based on the maximum cutoff distance, we create a set of \emph{tasks}, where some kinds tasks compute all the local interactions in a space voxel, and others compute pair-wise interactions between neighboring voxels. We create a thread queue where, at each time step, the worker threads grab the next available task, and locks the voxel or voxels that participate in this task. Thus, other ready threads grab the next available task, such that that task does not depend on any active spatial voxels. We have found this approach works well in ensuring all threads are busy and minimised thread contention. 

On the rendering side, we always render to an OpenGL rendering context. In desktop mode, we render to the on-screen rendering context, but in the server side, we use GLES to perform completely headless rendering, without using X11. In headless mode, we create an off-screen OpenGL frame buffer and render to this, and we copy the bits out of the frame buffer, compress it in real-time and generate a live-stream of image data that we display in an interactive web session using a custom Jupyter Widgets display widget. We currently can achieve about 10-15 frames per second when remote rendering, and we are working on improvements to this architecture which should increase frame rate to about 25 frames per second. 

The ability to run live, interactive simulations leveraging server compute power, and displaying the results to the user in a standard web browser represents a significant advance in broadening the impact of simulation technologies on the biomedical community at large. 

Mechanica is currently deployed and being used by our students on the nanoHUB development servers~(\url{https://nanohub.org/}, and we are currently working with the nanoHUB admin team to push this to production, such that it will be publicly accessible for anyone. 

We address the challenge of integrating a REPL loop, which enables users to evaluate expressions while the simulation is running through two different approaches. When Mechanica runs in desktop mode, in the command line \texttt{ipython} interactive environment, we integrate a \emph{display hook}. Here, the main ipython application runs a standard message loop required for any desktop application with a user interface, and we hook into this loop. On startup, we introspect the Python run-time, and inject our hook function into ipython's message loop. This, the simulation runs continuously, at the same time that the user's console is still accepting input.  When we run in a Jupyter notebook, we instead create a background thread that time-advances the simulation, renders each frame, and sends each rendered image over to the client's web browser to be displayed. This approach leaves does not block the Jupyter notebook, and enables the user to enter expressions in the input cells.

\section{Conclusion}

Lagrangian, particle based simulation approaches are the most widely used method in applications requiring high-performance, real-time physics and involve highly dynamics geometries and varying number of agents, such as in the computer games industries. Active-matter and biological matter simulations have far more in common with computer games than with engineered applications with fixed geometries, connectivity and agent number. Thus we have chosen to apply these techniques originally developed in the games industries to biological simulation. 

We have addressed one of the harder challenges in writing active matter and biological models: how to enable users to actually \emph{write} and develop models with a simple, easy to use modeling formalism. Our modeling API enables users to write complex biological models with very short, easy to read Python scripts. Models written with the Mechanica API are typically under 50 lines of standard Python. We leverage natural Python language level concepts such as classes, nested classes to enable users to write model types, and simply connect processes to type definitions. Because we use high level abstractions to represent physical concepts such as forces and processes, this allows a large separation between model specification, and numerical solver implementation. This large separation gives numerical solver developers a great deal of freedom in optimizing internal solver implementation without affecting end-user model development.  

We have demonstrated Mechanica's model specification API, and multi-process OpenGL accelerated rendering capabilities, where user models can run unchanged both in a desktop application, and server side in a Jupyter notebook. 

Mechanica is designed first and foremost as a self-contained \emph{library} with both Python and C APIs that make it easy to use Mechanica as a \emph{component} in a larger, well-established simulation platforms such as CompuCell3D or PhysiCell, where high-performance fluids, materials and agent simulation capabilities could be leveraged. 

\section{Future Plans}
One of the most important capabilities of a active-matter simulator is to enable local reaction networks at each agent. We are currently working on integrating the SBML JIT (Just in time) compiler and solver we have previously developed in the libRoadRunner~\cite{Somogyi:2015iz} library. However, local reaction networks time evolution can be extremely costly in terms of computational performance. Thus, we are working on porting our reaction kinetics solver code to use the GPU, and JIT compile reaction kinetics networks to SPIR-V (\url{https://www.khronos.org/registry/spir-v/}) to run on the GPU. 

\section{Availability}
Mechanica is released under a permissive Library-GPL license, which means that users of the library do not have to abide by any particular license. 

\begin{itemize}
    \item All source code is available at our GitHub repository at:\\ \url{https://github.com/AndySomogyi/mechanica}. 
    \item Complete documentation is available online at: \\
    \url{https://mechanica.readthedocs.io/}.
    \item Python PIP binaries are available for Mac and Windows, and we are in the processes of creating \texttt{manylinux} compatible binaries for Linux. PIP binaries are available at: \\ \url{https://pypi.org/project/mechanica/}
\end{itemize} 

\section{Acknowledgements}

This work was supported by National Institute of Health under grant number U24-EB028887-02. The content is solely the responsibility of the authors and does not necessarily represent the official views of the National Institutes of Health, the University of Washington or Indiana University, Bloomington.

\nolinenumbers

\bibliographystyle{plain}
\bibliography{papers}
\end{document}